\documentclass{PoS}

\title{
Recent developments on direct $\boldmath{CP}$ violation in the kaon system
and connection to $\boldmath{K \to \pi \nu \bar{\nu}}$ measurements}

\ShortTitle{Direct $CP$ violation in the kaon system
and connection to $K \to \pi \nu \bar{\nu}$}

\author{\speaker{Teppei Kitahara}
\\
{
Institute for Advanced Research, Nagoya University, Nagoya 464-8602, Japan,\\
Kobayashi-Maskawa Institute for the Origin of Particles and the
Universe, Nagoya University, Nagoya 464-8602, Japan,
 \\
and \\
Physics Department, 
Technion--Israel Institute of Technology, Haifa 3200003, Israel
}\\
        E-mail: \email{teppeik@kmi.nagoya-u.ac.jp}}

\abstract{
The first lattice result from the RBC and UKQCD Collaborations 
and improved perturbative calculations of $\varepsilon^{\prime}_K / \varepsilon_K$ have implied that the Standard-Model (SM) expectation deviates from measured values at the $2.8\,\sigma$ level. 
Since $\varepsilon^{\prime}_K / \varepsilon_K$ 
comes from $CP$-violating FCNC and is significantly suppressed in the SM, the discrepancy can be explained easily in several 
new physics (NP) models.
In this contribution, 
it is shown that correlations with the other rare decays, especially  
$K\to \pi \nu \overline{\nu}$ and $K_S \to \mu^+ \mu^-$, are crucial for discrimination of the NP models.
These channels 
can be probed precisely in the future 
by the NA62 and KOTO experiments for $K\to \pi \nu \overline{\nu}$ and LHCb experiment for  $K_S \to \mu^+ \mu^-$.
  } 

\FullConference{XIV International Conference on Heavy Quarks and Leptons (HQL2018)\\
		May 27- June 1, 2018\\
		Yamagata Terrsa, Yamagata,Japan}

\usepackage{amsmath}
\usepackage{cite}

\newcommand{\primed}{^{\prime}}
\def\beq#1\eeq{\begin{align}#1\end{align}}

\def\lsim{\mathrel{\rlap{\lower4pt\hbox{\hskip1pt$\sim$}}
    \raise1pt\hbox{$<$}}}                
\def\gsim{\mathrel{\rlap{\lower4pt\hbox{\hskip1pt$\sim$}}
    \raise1pt\hbox{$>$}}}  
\newcommand{\tev}{\,\mbox{TeV}}

\newcommand{\eq}[1]{Eq.~(\ref{#1})}

\newcommand{\imag}{\textrm{Im}\,}

\begin{document}

\boldmath  
\section{${\varepsilon^{\prime}_K}$ in the Standard Model}
\unboldmath  

Charge-parity ($CP$) violating flavour-changing neutral current (FCNC) decays of kaon are significantly suppressed by a small CKM component of
$\textrm{Im}[V_{ts}^{\ast} V_{td}] / |V_{us}^{\ast} V_{ud} |\sim 0.6 \times 10^{-3}$
 and a loop suppression factor in the Standard Model (SM), and hence are
extremely sensitive to new physics (NP).
Prime examples of such observables are
direct $CP$ violation in $K_L\to
\pi^+\pi^-,\,\pi^0\pi^0$ decays, the branching fraction of $K_L\to\pi^{0}\nu\overline{\nu}$, and the flavour-tagged
asymmetry in $K_S \to \mu^+ \mu^-$ decay.

In $K_L \to \pi\pi$ decays, 
one can distinguish between two types of  $CP$
violation: 
direct ($\varepsilon_{K}^{\prime}$) and indirect $CP$ violation ($\varepsilon_K$).
Both kinds of $CP$ violation have been precisely measured by many kaon experiments.
Note that 
$\varepsilon_{K}^{\prime}$ is smaller than $\varepsilon_K$ by three orders of magnitude.
  This strong suppression comes from the smallness of the $\Delta I = 3/2$ decay (to $I=2$ state) compared to 
the  $\Delta I = 1/2$ decay (to $I=0$ state), namely the $\Delta I = 1/2$ rule,  and an
accidental cancellation of leading penguin contributions in the SM.
Their suppressions lead to high sensitivity to NP.
Until recently, large theoretical uncertainties precluded reliable
predictions for $ \varepsilon _K^{\prime}$. 
Although SM predictions of $ \varepsilon _K^{\prime}$ using chiral  perturbation theory (ChPT) are consistent with the experimental value, their theoretical uncertainties are large.
In contrast, a calculation by the dual QCD approach \cite{Buras:1985yx,Buras:2014maa}
 finds the SM value much below the experimental one.
  A major breakthrough has been obtained from the recent lattice-QCD calculations of the hadronic matrix elements by the RBC-UKQCD collaboration \cite{Blum:2015ywa, Bai:2015nea}, which supports  
the latter  result.

\begin{figure}[t]
  \begin{center} 
  \hspace{-0.6cm}
    \includegraphics[width=1. \textwidth, 
    bb
     = 0 0 351 206]{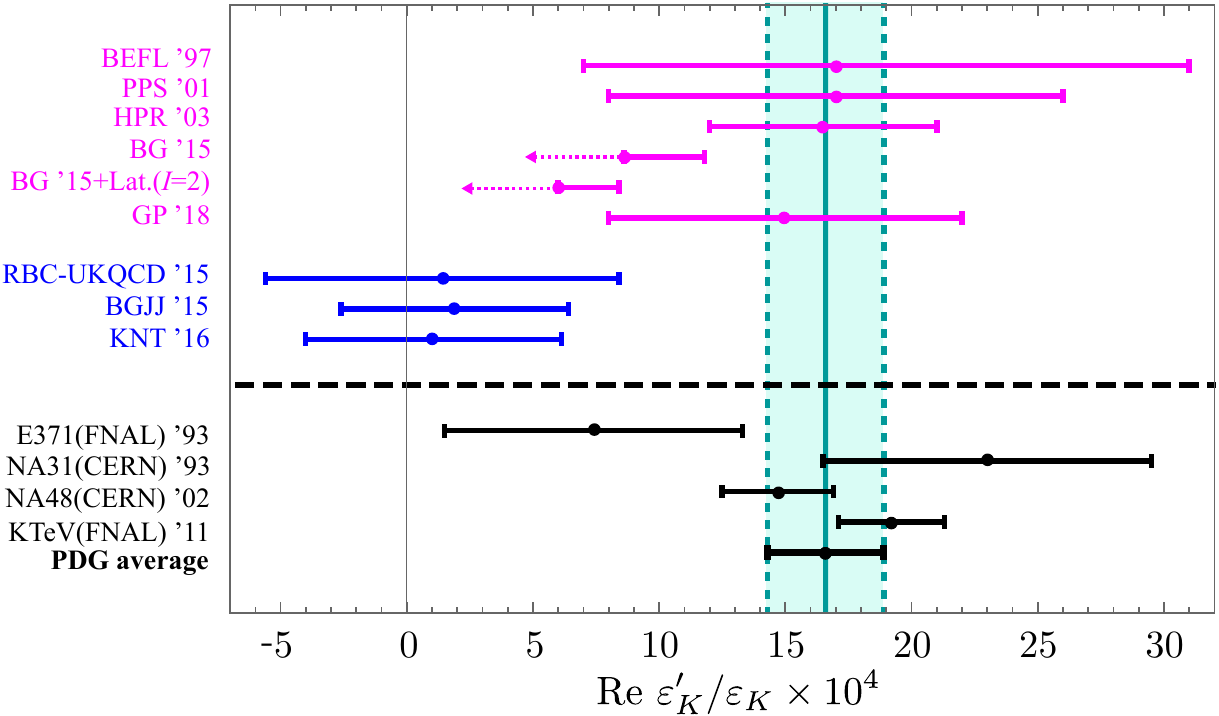}
    \caption{
    Compilation of representative SM predictions and the experimental values for Re$(\varepsilon'_K / \varepsilon_K)$. 
     All error bars represent 1\,$\sigma$ range. 
     The SM predictions are taken from Bertolini \emph{et al.} (BEFL '97) \cite{Bertolini:1997nf},  Pallante \emph{et al.} (PPS '01) \cite{Pallante:2001he}, Hambye \emph{et al.} (HPR '03)~\cite{Hambye:2003cy}, Buras and G\'erard (BG '15) \cite{Buras:2015xba, Buras:2016fys},  Gisbert and Pich (GP '18) \cite{Gisbert:2017vvj},  RBC-UKQCD lattice result \cite{Blum:2015ywa, Bai:2015nea}, Buras \emph{et al.} (BGJJ '15) \cite{Buras:2015yba}, and Kitahara \emph{et al.} (KNT '16)  \cite{Kitahara:2016nld},
      where magenta bars are based on analytic approaches to hadronic matrix elements while blue bars are based on lattice results.
    The thick black one is the world average of the experimental values \cite{Patrignani:2016xqp}.   }
\label{fig:status}
\end{center}
\end{figure}

A compilation of representative SM predictions and the experimental values for Re$(\varepsilon'_K / \varepsilon_K)$ is given in Fig.\,\ref{fig:status}.
The SM predictions (magenta and blue bars)
 are taken from Refs.~\cite{Bertolini:1997nf, Pallante:2001he, Hambye:2003cy, Buras:2015xba, Buras:2016fys, Gisbert:2017vvj, Blum:2015ywa, Bai:2015nea, Buras:2015yba, Kitahara:2016nld}.
The experimental values (black bars) are taken from Refs.~\cite{Gibbons:1993zq, Barr:1993rx, Batley:2002gn, Abouzaid:2010ny}.
The thick black one is the world average of data \cite{Patrignani:2016xqp}
\beq
   \textrm{Re}\left( \varepsilon_K^\prime / \varepsilon_K \right)_{\rm exp} =
\left(16.6 \pm 2.3 \right) \times 10^{-4}.
 \label{PDG}
\eeq

In order to predict $\varepsilon_K^\prime$ in the SM, 
one has to calculate the hadronic matrix elements of four-quark operators using nonperturbative methods.
The magenta bars  in Fig.\,\ref{fig:status}  have utilized analytic approaches  to calculate  them: 
 chiral quark model (BEFL '97),   ChPT (PPS '01 and GP '18) with  the minimal hadronic approximation (HPR '03), and the dual   QCD approach (BG '15). 
 On the other hand, 
 a determination of all hadronic matrix elements from lattice QCD has 
  been obtained by the RBC-UKQCD collaboration \cite{Blum:2015ywa, Bai:2015nea}, and 
 the blue bars are based on the lattice result:
\beq%
\varepsilon_K^\prime / \varepsilon_K=
\begin{cases}
  \left(1.9 \pm 4.5  \right) \times 10^{-4}&(\textrm{BGJJ~'15}),\\
 {\left(1.06 \pm  5.07 \right)}\times 10^{-4}&(\textrm{KNT~'16}).
  \end{cases}
  \label{discrepancy}
  \eeq %
  These results are obtained by next-to-leading order (NLO) calculations exploiting 
    $CP$-conserving data to reduce hadronic uncertainties and include isospin-violating contributions \cite{Cirigliano:2003nn} which are not included in the lattice result.  
    Furthermore, the latter result contains an additional $\mathcal{O}(\alpha_{EM}^2/\alpha^2_s)$ correction, which appears only in this order, and also
  utilizes a new analytic solution of the renormalization group (RG) equation which avoids the  problem of singularities in the NLO terms.
   The two
    numbers in~\eq{discrepancy} disagree with the experimental
value in \eq{PDG}  by $2.9\,\sigma$ \cite{Buras:2015yba} and $2.8\,\sigma$~{\cite{Kitahara:2016nld}}, respectively. 
   The  uncertainties are  dominated by the lattice statistical and systematic uncertainties for the $I = 0$ amplitude.

The main difference between the analytic approach  and the lattice result  comes from a hadronic matrix element   
$ \langle ( \pi \pi )_{I=0} \left| Q_6 \right| K^0 \rangle \propto B_6^{(1/2)} $  which controls the largest  positive  contribution to $\varepsilon'_K/\varepsilon_K$ $[ Q_6 =  ( \bar{s}_{\alpha} d_{\beta} )_{V-A} \sum_q ( \bar{q}_{\beta} q_{\alpha} )_{V+A}]$.
In ChPT, a large value has been obtained: $B_6^{(1/2)} \sim 1.6 $ (BEFL '97), $\sim1.6$ (PPS '01), and $\sim3$ (HPR '03, see Ref.~\cite{Buras:2015xba}).
On the other hand, the dual QCD approach predicts a small  number, $B_6^{(1/2)} \leq  B_8^{(3/2)} \sim  0.8$ (BG '15).
The current lattice result is consistent with the latter result: $B_6^{(1/2)}  = 0.56 \pm 0.20$ \cite{Bai:2015nea, Kitahara:2016nld}.

Although the lattice simulation \cite{Bai:2015nea}
includes final-state interactions partially along the line of Ref.~\cite{Lellouch:2000pv}, 
the lattice result of the strong phase
shift $\delta_{0}$  is smaller than the phenomenological expectation at $2.8\,\sigma$ level \cite{Colangelo:2015kha}.
Meanwhile, the phase shift $\delta_2$ of the lattice result is consistent with the phenomenological expectation.
Also, the lattice result explains the $\Delta I = 1/2$ rule for the first time at $1\,\sigma$ level \cite{Blum:2015ywa,Bai:2015nea, Buras:2016fys},
\beq
(\textrm{Re} A_0 / \textrm{Re} A_2)_{\textrm{exp}}  = 22.45 \pm  0.05, ~~~ (\textrm{Re} A_0 / \textrm{Re} A_2)_{\textrm{Lat.}}  = 31.0 \pm  11.1.
\eeq 
In the near future, the increasing precision of lattice calculations 
using more sophisticated methods will further sharpen the SM
  predictions in \eq{discrepancy} and  answer the question about  NP in $\varepsilon_K^\prime/\varepsilon_K$.
 In the lattice preliminary result, the discrepancy of the  strong phase
shift $\delta_{0}$ is resolved~\cite{lattice_prospect}.

Several NP models including supersymmetry (SUSY) can explain the discrepancy of $ \varepsilon'_K / \varepsilon_K$.
 It is known that such NP models are likely to predict deviations of the other rare decay   branching ratios from the SM predictions, 
 especially $K\to \pi \nu \overline{\nu}$ which includes  $CP$-violating FCNC decay and can be probed precisely in the near future by the NA62 and KOTO experiments.
In this contribution, 
based on the lattice result of $ \varepsilon'_K / \varepsilon_K$ and Eq.~\eqref{discrepancy},
we present 
correlations between $\varepsilon^{\prime}_K / \varepsilon_K$ and $\mathcal{B}(K\to \pi \nu \overline{\nu})$ 
in two types of NP scenarios: a box dominated scenario and a $Z$-penguin dominated one, and discuss how to distinguish between them.

 \section{Box dominated scenario}
 
 We first focus on the box dominated scenario, where all NP contributions to $|\Delta S|=1 $ and $ |\Delta S| = 2$ processes are dominated by four-fermion box diagrams.
Such a situation is realized in the minimal supersymmetric standard model (MSSM)~\cite{Kitahara:2016otd}.
The desired effect in $\varepsilon_K\primed$ is generated via gluino-squark box diagrams when a mass difference between the right-handed up and down squarks exists \cite{Kagan:1999iq,Grossman:1999av}.

While sizable effects in $\varepsilon_K\primed$ are obtained by the gluino box  contributions,
 simultaneous efficient suppression of the SUSY QCD contributions to $\varepsilon_K$ can also be achieved. 
 The Majorana nature of the gluino leads to a suppression of $|\Delta S | = 2 $ gluino  box contributions to $\varepsilon_K$, where there are two such diagrams (crossed and  uncrossed boxes) with opposite signs. 
If the gluino mass $m_{\tilde g}$ equals roughly 1.5 times the average down squark 
mass $M_S$,  both contributions to $\varepsilon_K$ cancel \cite{Crivellin:2010ys}. 
Note that this suppression appears only when a hierarchy $\Delta_{Q,12} \gg \Delta_{\bar{D},12}$ or  $\Delta_{Q,12} \ll  \Delta_{\bar{D},12}$ is satisfied, where the following notation is used for the squark mass matrices: $ M^2_{X, ij} = m^2_{X} \left( \delta_{ij} + \Delta_{X, ij} \right), $
with $X = Q,~\bar{U}$,~or~$\bar{D}$.

\boldmath  
\subsection{Contributions to $\varepsilon_K^\prime$ }\unboldmath

The master equation for $\varepsilon_K^\prime / \varepsilon_K$ (see e.g., Ref.~\cite{Buras:2015yba}) reads:
\beq%
\frac{\varepsilon_K^\prime}{\varepsilon_K} = \frac{\omega_{+}}{\sqrt{2}
  {|}\varepsilon_K^{\textrm{exp}}{|} \textrm{Re} A_0^{\textrm{exp}} } \left[
  \frac{\textrm{Im} A_2 }{\omega_{+}} - \left( 1-
    \hat{\Omega}_{\textrm{eff}} \right) \textrm{Im} A_0 \right],
\label{eq:mas}
\eeq%
with $\hat{\Omega}_{\textrm{eff}} = (14.8\pm
8.0)\times 10^{-2}$,  {the measured
  $|\varepsilon_K^{\rm exp}|$,} $\omega_{+}= (4.53 \pm 0.02)\times10^{-2}$,  and the amplitudes $A_I = \langle (\pi \pi)_I |
\mathcal{H}^{\left|\Delta S\right| = 1} | K^0 \rangle$ involving the
effective $|\Delta S|=1$ {Hamiltonian} $ \mathcal{H}^{\left|\Delta
    S\right|}$. $I=0,2$ represents the strong isospin of the final two-pion
state. 
  The gluino box diagrams contribute to $\imag A_2$ when $m_{\bar{U}}\neq m_{\bar{D}}$. Because these contributions are  governed by the strong interaction and there is  an enhancement factor $1/\omega_+ = 22.1$ for the $\imag A_2$ term in \eqref{eq:mas}, they easily become the largest contribution to $\varepsilon'_K / \varepsilon_K$.
 To obtain the desired large effect in $\varepsilon_K^\prime$, 
 the flavour mixing has to be in the left-handed squark mass matrix.  
 The opposite situation with right-handed flavour mixing and $\tilde u_L$-$\tilde d_L$ mass splitting is not possible because of the $\mathrm{SU}(2)_L$ invariance.

\begin{figure}[t]
\begin{center}
    \includegraphics[width=0.49 \textwidth, 
    bb
     = 0 0 360 362]{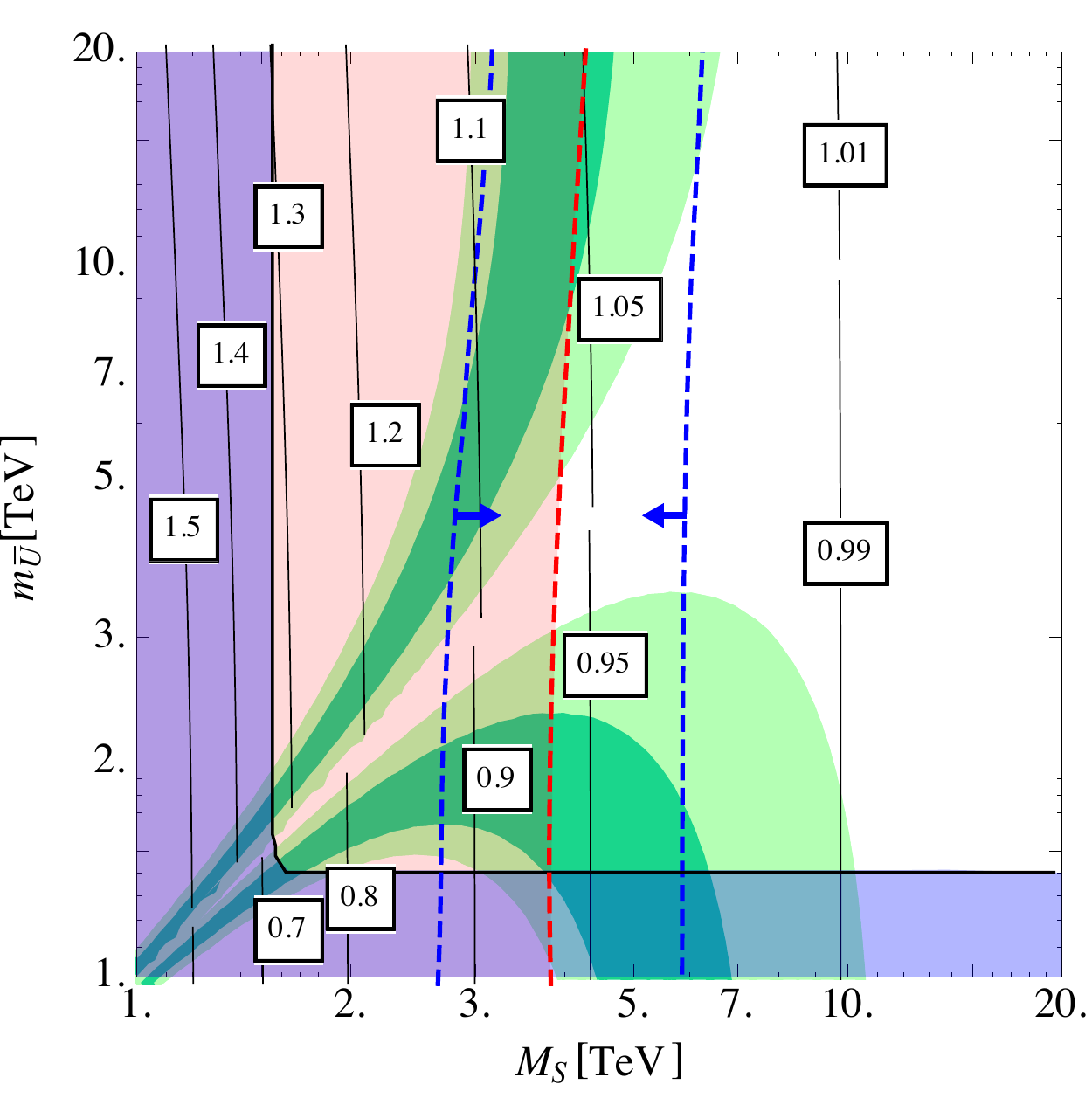} 
\includegraphics[width=0.49\textwidth, bb= 0 0 352 345]{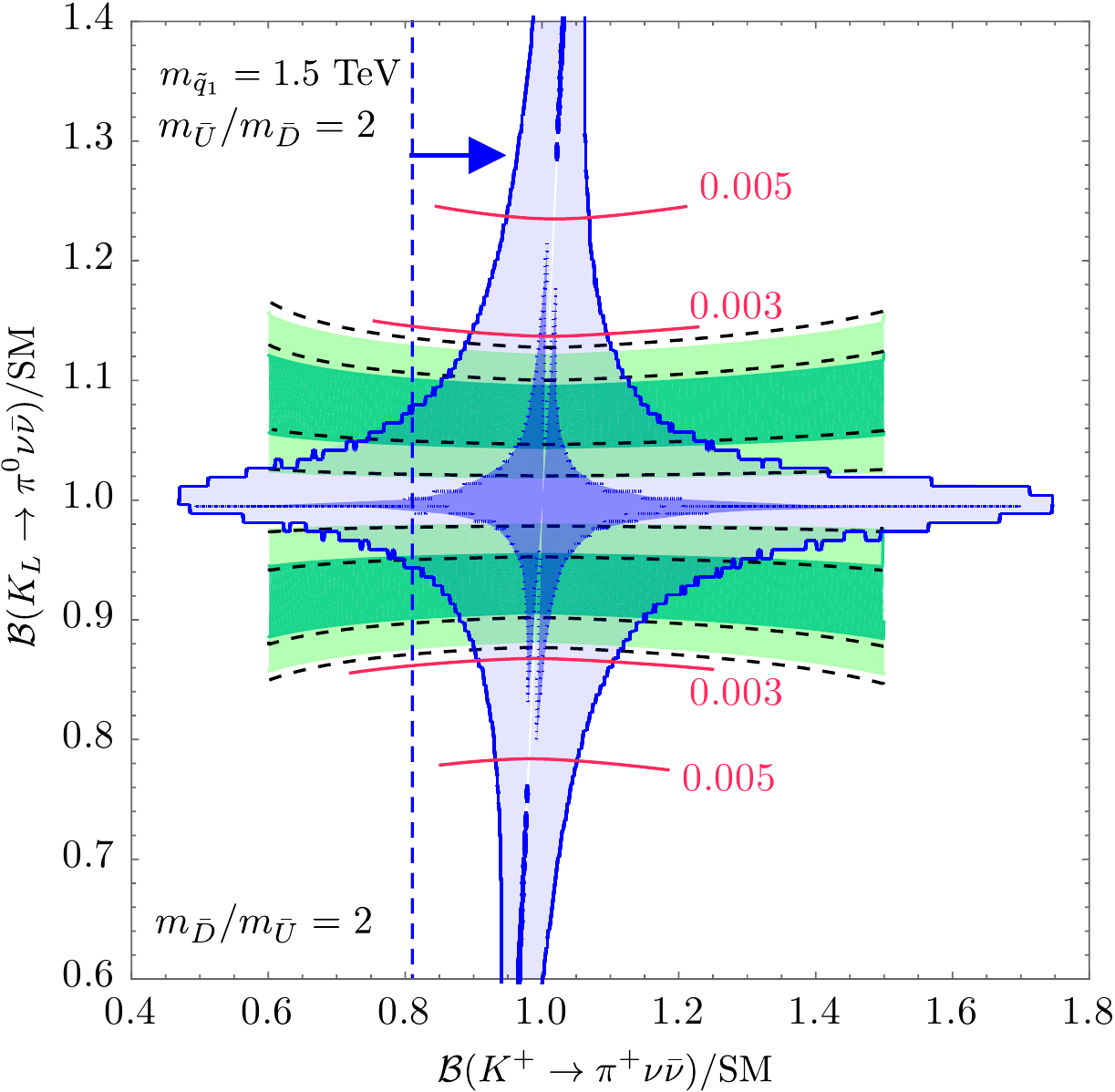}
\caption[]{
In the left panel, parameter constraints from $\varepsilon_K$ and the LHC results are shown by the red and blue regions.
The correlation with $\mathcal{B}(K \to
\pi \nu\overline{\nu})$ is also shown in the right panel.
The $\varepsilon_K^\prime/\varepsilon_K$
  discrepancy is resolved at the $1\,\sigma$\,($2\,\sigma$) level within
  the dark (light) green region in both panels.  
The light (dark) blue region requires a milder parameter
  tuning than 1\,\%\,(10\,\%) of the gluino mass and  the $CP$ violating phase in order to suppress contributions to $\varepsilon_K$. 
   The red contour represents the SUSY contributions to
  $\varepsilon^{\prime}_K / \varepsilon_K$.
 }
\label{fig:boxresult}
\end{center}
 \end{figure}

In the left panel of Fig.~\ref{fig:boxresult},
the portion of the squark mass plane which simultaneously explains
$\varepsilon_K^\prime/\varepsilon_K$ discrepancy and $\varepsilon_K$ constraint is shown.    
As input, we take the grand-unified theory (GUT) relation for gaugino masses, 
$m_{\tilde{g}}/M_S = 1.5 $ for the suppressed $\varepsilon_K$, and $m_{Q} = m_{\bar{D}} = \mu_{\rm SUSY}=M_S$ with varying $m_{\bar U}$.
The universal slepton mass is set to be $m_{L} = 300$ GeV.
Furthermore, the trilinear SUSY-breaking matrices $A_q$ are set to zero, $\tan \beta =10$, and the only nonzero off-diagonal element of the squark mass  matrices is $\Delta_{Q, 12} = 0.1 \exp(- i \pi/4)$  {for the left-handed squark sectors} for $m_{\bar U} > m_{\bar D} =  M_S$
(upper branch) and $\Delta_{Q, 12} = 0.1 \exp( i 3 \pi/4)$ for $m_{\bar U} < m_{\bar D} = M_S$ (lower branch).

\boldmath  
\subsection{$\mathcal{B}(K_L{\to\pi^{0}
  \nu\overline{\nu}})$ and $\mathcal{B}(K^+{\to\pi^{+}
  \nu\overline{\nu}})$ }\unboldmath

The SUSY contributions to $ \varepsilon_K$ can be suppressed by the crossed and  uncrossed box diagrams when the gluino mass is heavier than the squark mass, while there is no such cancellation in a chargino box
contribution to $K_L\to \pi^0 \nu \overline{\nu}$ and $K^+\to \pi^+ \nu
\overline{\nu}$ which permits potentially large effects.
We investigate the correlation between $\varepsilon_K\primed$ and $\mathcal{B}(K \to \pi \nu \overline{\nu})$ varying the following parameters:
\begin{equation}
|\Delta_{Q,12}|,~\theta,~M_3,~m_{\bar{U}}/m_{\bar{D}}, 
\end{equation}
 with $0 < |\Delta_{Q,12}| < 1$ and $0 < \theta < 2 \pi$. 
 Here, defining the bilinear terms for the squarks as 
  $\theta \equiv \textrm{arg}(\Delta_{Q,12})$.  
We fix the slepton mass and the lightest squark mass close to the experimental limit ($m_L
= 300\,$GeV and $m_{\tilde{q}_1} = 1.5\,$TeV) and use GUT
relations among all three gaugino masses.

The  right panel of Fig.~\ref{fig:boxresult} shows the correlations between    $\varepsilon_K\primed$ and $\mathcal{B}(K \to \pi \nu \overline{\nu})$
  in the $\mathcal{B}(K_L\to\pi^{0}
\nu\overline{\nu})$--$\mathcal{B}(K^{+}\to\pi^{+} \nu\overline{\nu})$
plane which is normalized by their SM predictions \cite{Crivellin:2017gks}.
We find that  the necessary amount of the tuning in the
gluino mass and the $CP$ violating phase   in order to suppress contributions to $\varepsilon_K$  determines deviations of $\mathcal{B}(K \to\pi \nu\overline{\nu})$ from the SM values.
A quantity which parameterizes the fine-tuning parameter is defined in Ref.~\cite{Crivellin:2017gks}.
The current
$\varepsilon_K^\prime/\varepsilon_K$ discrepancy  between \eq{PDG} and  \eq{discrepancy}  is resolved at
$1\,\sigma$\,($2\,\sigma$) within the dark (light) green region.  
We used
$m_{\bar{D}}/m_{\bar{U}} = 2$ with $m_{\bar{U}} = m_Q$ for $0 <
\theta < \pi$, and $m_{\bar{U}}/m_{\bar{D}} = 2$  with $m_{\bar{D}}
= m_Q$ for $\pi < \theta < 2 \pi$.  
Numerically, we observe $\mathcal{B}(K_L \to \pi^0\nu\overline{\nu})/\mathcal{B}^{\rm SM}
  (K_L \to \pi^0\nu\overline{\nu})\lesssim 2\,(1.2)$ and $\mathcal{B}(K^+ \to
  \pi^+\nu\overline{\nu})/\mathcal{B}^{\rm SM}(K^+ \to \pi^+\nu\overline{\nu})
  \lesssim 1.4\,(1.1)$ in light of $\varepsilon^{\prime}_K / \varepsilon_K$ discrepancy, 
  if all squarks are heavier than $1.5\,\tev$ and if a $1\,(10)\,\%$ fine-tuning is permitted.

We also observe a strict correlation between $\mathcal{B}(K_L \to
\pi^0\nu\overline{\nu})$ and $m_{\bar{U}}/m_{\bar{D}}$: $\mbox{sgn}\,
[\mathcal{B}(K_L \to \pi^0\nu\overline{\nu})-\mathcal{B}^{\rm SM} (K_L
\to \pi^0\nu\overline{\nu}) ] = \mbox{sgn}\,[m_{\bar{U}}-m_{\bar{D}}]$.
Thus, $\mathcal{B}(K_L \to \pi^0\nu\overline{\nu})$ 
can indirectly determine  whether  the right-handed up or  down squark is the heavier one.

\section{$Z$-penguin dominated (modified $Z$-coupling) scenario}

Next we focus on the $Z$-penguin dominated scenario. 
The largest negative contribution to $\varepsilon^{\prime}_K $ comes from 
 $Z$-penguin diagrams in the SM.
 Since in the SM there is a large numerical cancelation between QCD- and  $Z$-penguin contributions to $\varepsilon^{\prime}_K / \varepsilon_K$, 
 a modified $Z$ flavour-changing ($s$--$d$) interaction from NP can explain the current $\varepsilon^{\prime}_K / \varepsilon_K$ easily \cite{Buras:2014sba,Buras:2015yca}.
Then,  
the decay, $s\to d\nu  \overline{\nu}$, proceeding   through an intermediate $Z$ boson, must be modified by  the NP.
Therefore, the branching ratios of $K \to \pi\nu\bar\nu$ are likely to deviate from the SM predictions once the $\varepsilon^{\prime}_K/\varepsilon_K$ discrepancy  between \eq{PDG} and  \eq{discrepancy}  is explained by the modified $Z$-coupling. 
They could be a signal to test the scenario. 
In the MSSM, such a scenario is also realized when the off-diagonal components of the  trilinear SUSY-breaking couplings are large   \cite{Tanimoto:2016yfy, Endo:2016aws, Endo:2017ums}.

Such a signal is constrained by the $\varepsilon_K$.
The modified $Z$ couplings affect the $\varepsilon_K$  via the so-called double penguin diagrams. 
Such a contribution is enhanced when there are both left-handed  and right-handed couplings because of the chiral enhancement of the hadronic matrix elements. 
The important point is that 
since the left-handed coupling is already present in the SM, 
the right-handed coupling must be constrained even without NP contributions to the left-handed one. 
Such interference contributions between the NP and the SM have been overlooked in the literature. 
References \cite{Endo:2016tnu, Bobeth:2017xry, Endo:2017ums} have revisited the modified $Z$-coupling scenario including the interference contributions using a framework of the SMEFT, and found 
the parameter regions allowed by the indirect $CP$ violation change significantly.

We find that similar to the previous section, the deviations of $\mathcal{B}(K \to \pi \nu \overline{\nu})$ from the SM values are  determined  by the necessary amount of the tuning in NP contributions  to $\varepsilon_K$.
We parametrize it by $\xi$:  A degree of the NP parameter tuning is represented  by $1 / \xi$, 
e.g., $\xi = 10$ means that the model parameters are tuned at the 10\% level. The definition of $\xi$ is given in Ref.~\cite{Endo:2016tnu}.

\begin{figure}
\begin{center}
\includegraphics[width=0.49\textwidth, bb = 0 0 561 472]{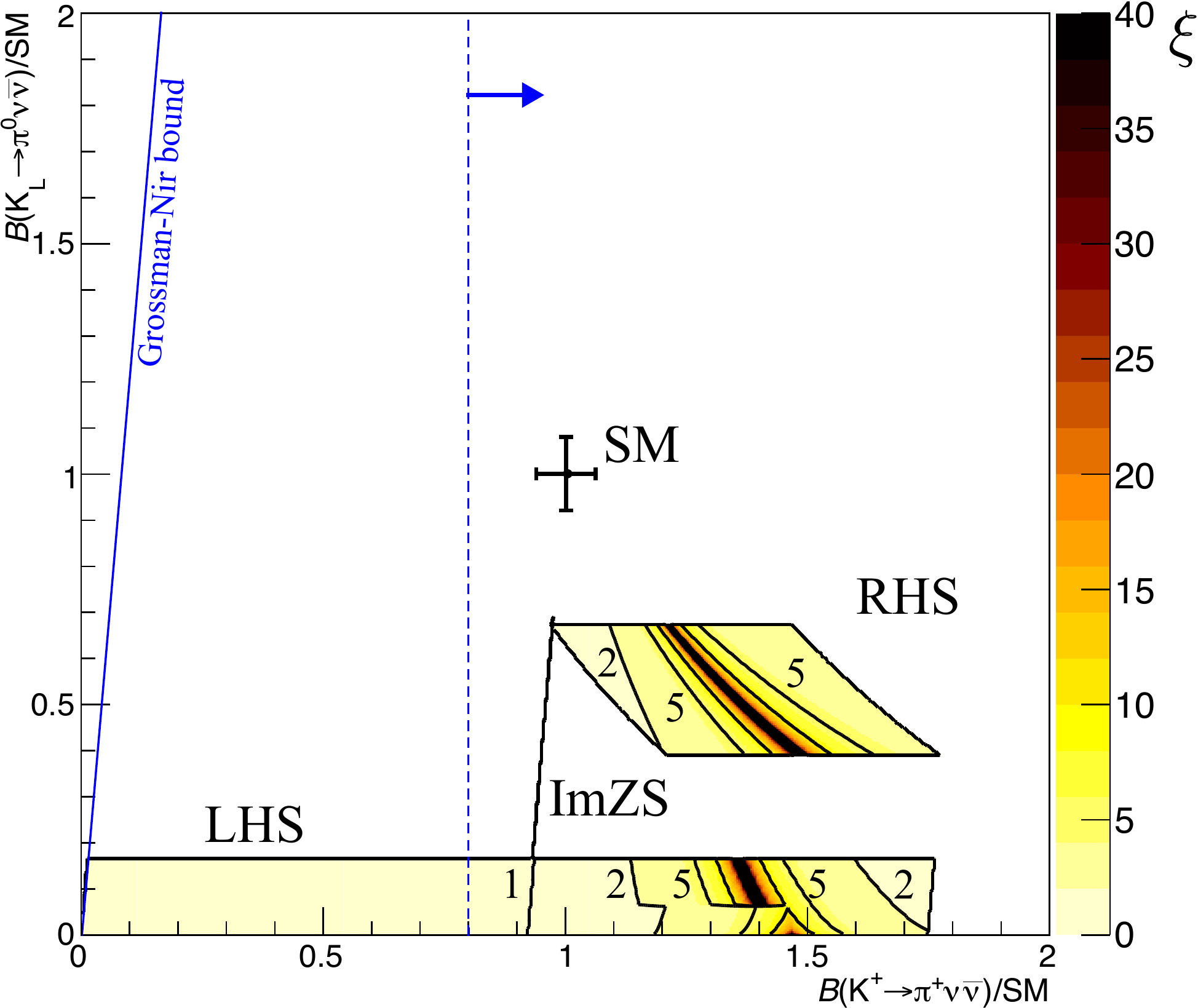} 
\includegraphics[width=0.49\textwidth, bb = 0 0 558 472]{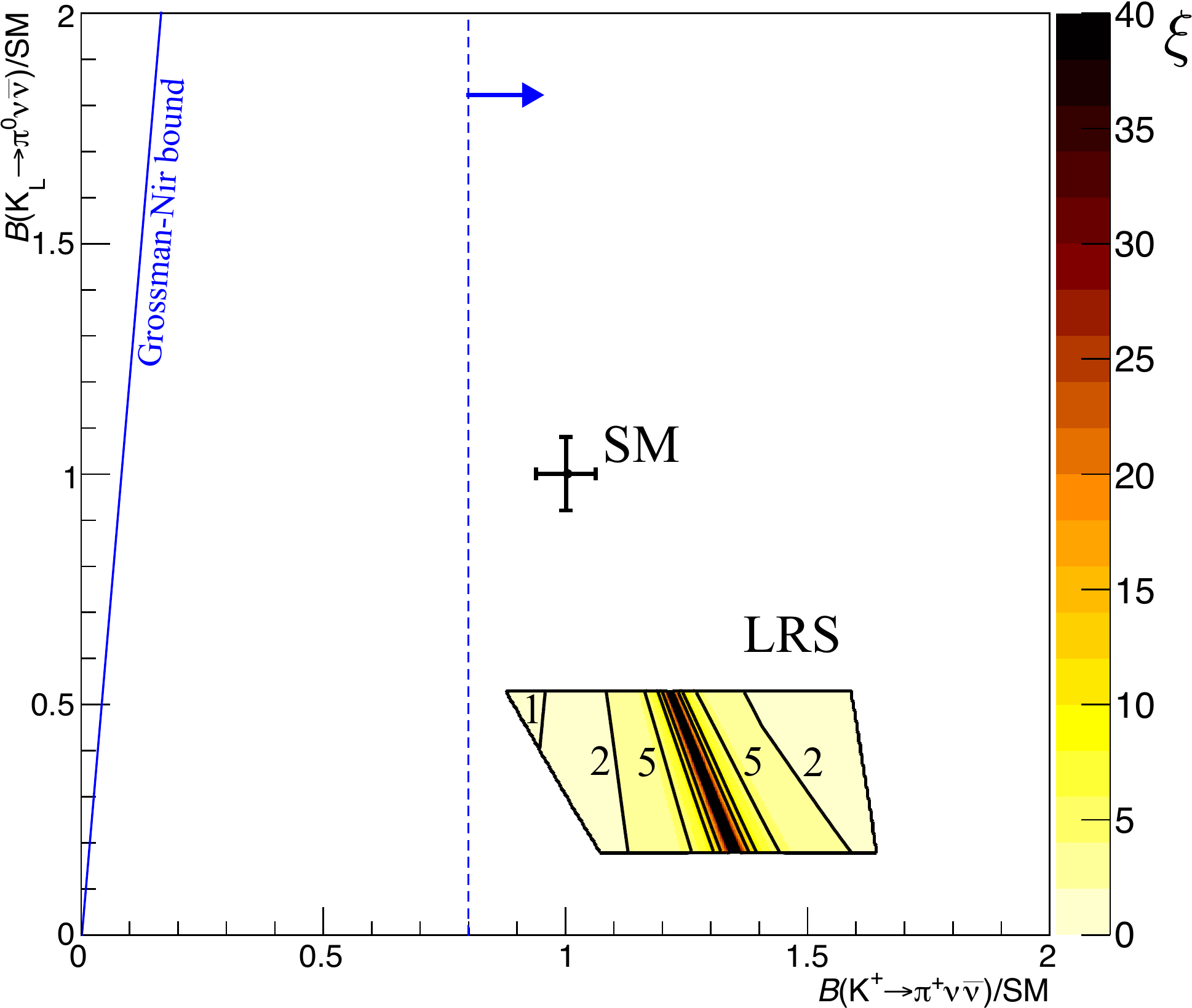}
\caption[]{
Contours of the tuning parameter $\xi$ are shown in the simplified modified $Z$-coupling scenarios: LHS, RHS, and ImZS (left panel) and LRS (right).
In the colored regions, $\varepsilon^{\prime}_K/ \varepsilon_K$ is explained at $1\sigma$.
}
\vspace{-0.4cm}
\label{fig:simplified}
\end{center}
\end{figure}

In Fig.~\ref{fig:simplified}, contours of the tuning parameter $\xi$ are shown for the simplified scenarios:  LHS (all NP effects  appear as left-handed), RHS (all NP effects  appear as right-handed), ImZS (NP effects are purely imaginary), and LRS (left-right symmetric scenario) on the plane of the branching ratios of $K \to \pi \nu \overline{\nu}$ which are normalized by their SM predictions.
We scanned the whole parameter space of the modified $Z$-coupling in each scenario, and selected the parameters where $\varepsilon^{\prime}_K / \varepsilon_K$ is explained at the $1\sigma$ level.
The experimental bounds from $\varepsilon_K$, $\Delta M_K$, and $\mathcal{B}(K_L \to \mu^{+} \mu^{-})$ are satisfied. 
In most of the allowed parameter regions, $\xi = \mathcal{O}(1)$ is obtained.
Thus, one does not require tight tunings in these simplified scenarios. 
In the figures, 
$\mathcal{B}(K_{L}\to \pi^{0} \nu \overline{\nu} )$ is always smaller than the SM value by more than 30\%. 
In LHS, 
$\mathcal{B}(K_{L}\to \pi^{0} \nu \overline{\nu} )$ is much smaller and could be ruled out if a SM-like  value  is measured.
On the other hand, $\mathcal{B}(K^{+}\to \pi^{+}\nu \overline{\nu})$ depends on the scenarios. 
In LHS, we obtain $0 < \mathcal{B}(K^{+}\to \pi^{+}\nu \overline{\nu})/ \mathcal{B}(K^{+}\to \pi^{+}\nu \overline{\nu})_{\rm SM} < 1.8$. 
In RHS, $\mathcal{B}(K^{+}\to \pi^{+}\nu \overline{\nu})$ is comparable to or larger than the SM value, but cannot be twice as large.
In ImZS, the branching ratios are perfectly correlated and $\mathcal{B}(K^{+}\to \pi^{+}\nu \overline{\nu})$ does not deviate from the SM one.  
In LRS, $\mathcal{B}(K_{L}\to \pi^{0} \nu \overline{\nu} )$ does not exceed about a half of the SM value.
The more general situation is discussed in Ref.~\cite{Endo:2016tnu}.

\section{Discussion and conclusions}

In this talk, we presented the current situation for $\varepsilon^{\prime}_K / \varepsilon_K$ within the SM.
The first lattice result and the improved perturbative calculations have shown 
the discrepancy between the predicted value and the data.
Several NP models can explain the discrepancy of $ \varepsilon'_K / \varepsilon_K$, and then $\mathcal{B}(K\to \pi \nu \overline{\nu})$ are predicted to deviate from the SM predictions.
We have shown the correlations between $\varepsilon^{\prime}_K / \varepsilon_K$, 
$\mathcal{B}(K_L\to \pi^0 \nu \overline{\nu})$, and $\mathcal{B}(K^+\to \pi^+ \nu \overline{\nu})$   
in two different NP scenarios; the box dominated scenario and the $Z$-penguin dominated one.
 It is found that  
the constraint from $\varepsilon_K$  produces distinguishable correlations.
In the future,  measurements of $\mathcal{B}(K\to \pi \nu \overline{\nu})$ will be significantly improved. 
The NA62 experiment at CERN measuring $\mathcal{B}(K^+\to \pi^+ \nu \overline{\nu})$
 is aiming to reach a precision of 10\,\% compared to the SM value \cite{NA62:2017rwk}.  
Concerning $K_L \to \pi^0 \nu \overline{\nu}$, the KOTO experiment at J-PARC is aiming in a first step
to measure  $\mathcal{B}(K_L \to \pi^0 \nu \overline{\nu})$ around the SM sensitivity.
Furthermore, the KOTO-step2 experiment will aim at 100 events for the SM branching ratio, implying
a precision of 10\,\% of this measurement.
Therefore, 
we conclude that when the $\varepsilon^{\prime}_K / \varepsilon_K$ discrepancy is explained by a NP contribution, the NA62 experiment could probe whether a modified $Z$-coupling scenario is realized or not, and KOTO-step2 experiment can distinguish  the box dominated scenario and the simplified modified $Z$-coupling scenario.

We should comment on $K_S \to \mu^+ \mu^-$ decay which 
 proceeds via long-distance $CP$-conserving and short-distance $CP$-violating 
processes. 
Since the decay rate is dominated by the former, whose uncertainty is
large, the sensitivity to the short-distance contributions is
diminished. 
However, it is pointed out that the short-distance contribution is significantly amplified through interference
between the $K_L$ and $K_S$ states in the neutral kaon beam~\cite{DAmbrosio:2017klp}.
Therefore, one can also distinguish the NP scenarios using  the correlation with $K_S \to \mu^+ \mu^-$.
Such a correlation has been investigated in the box dominated scenario (with large $\tan \beta$) \cite{ Chobanova:2017rkj} and the modified $Z$-coupling scenario \cite{Endo:2017ums,DAmbrosio:2017klp}.

\section*{Acknowledgment}
I would like to thank   
Veronika Chobanova,
Andreas Crivellin, 
Giancarlo D'Ambrosio, 
Motoi Endo, 
Toru Goto,
Miriam Lucio Martinez,
  Diego Martinez Santos,
Satoshi Mishima,
 Ulrich Nierste, 
  Isabel Suarez Fernandez,
 Paul Tremper, 
  Daiki Ueda,
   and Kei Yamamoto for fruitful collaborations on  the presented work.
I also want to warmly thank the organizers of Heavy Quarks and Leptons 2018
for inviting and giving me the opportunity to present these results in a great conference.

\bibliographystyle{JHEP}
\bibliography{bib}

\providecommand{\href}[2]{#2}\begingroup\raggedright\begin{thebibliography}{10}

\bibitem{Buras:1985yx}
A.~J. Buras and J.~M. G\'erard, \emph{{$1/N$ Expansion for Kaons}},
  \href{https://doi.org/10.1016/0550-3213(86)90489-X}{\emph{Nucl. Phys.}
  {\bfseries B264} (1986) 371}.

\bibitem{Buras:2014maa}
A.~J. Buras, J.-M. G\'erard and W.~A. Bardeen, \emph{{Large $N$ Approach to
  Kaon Decays and Mixing 28 Years Later: $\Delta I = 1/2$ Rule, $\hat B_K$ and
  $\Delta M_K$}},
  \href{https://doi.org/10.1140/epjc/s10052-014-2871-x}{\emph{Eur. Phys. J.}
  {\bfseries C74} (2014) 2871}
  [\href{https://arxiv.org/abs/1401.1385}{{\ttfamily 1401.1385}}].

\bibitem{Blum:2015ywa}
T.~Blum et~al., \emph{{$K \rightarrow \pi\pi$ $\Delta I=3/2$ decay amplitude in
  the continuum limit}},
  \href{https://doi.org/10.1103/PhysRevD.91.074502}{\emph{Phys. Rev.}
  {\bfseries D91} (2015) 074502}
  [\href{https://arxiv.org/abs/1502.00263}{{\ttfamily 1502.00263}}].

\bibitem{Bai:2015nea}
{\scshape RBC, UKQCD} collaboration, Z.~Bai et~al., \emph{{Standard Model
  Prediction for Direct CP Violation in K→ππ Decay}},
  \href{https://doi.org/10.1103/PhysRevLett.115.212001}{\emph{Phys. Rev. Lett.}
  {\bfseries 115} (2015) 212001}
  [\href{https://arxiv.org/abs/1505.07863}{{\ttfamily 1505.07863}}].

\bibitem{Bertolini:1997nf}
S.~Bertolini, J.~O. Eeg, M.~Fabbrichesi and E.~I. Lashin, \emph{{Epsilon-prime
  / epsilon at O(p**4) in the chiral expansion}},
  \href{https://doi.org/10.1016/S0550-3213(97)00786-4}{\emph{Nucl. Phys.}
  {\bfseries B514} (1998) 93}
  [\href{https://arxiv.org/abs/hep-ph/9706260}{{\ttfamily hep-ph/9706260}}].

\bibitem{Pallante:2001he}
E.~Pallante, A.~Pich and I.~Scimemi, \emph{{The Standard model prediction for
  epsilon-prime / epsilon}},
  \href{https://doi.org/10.1016/S0550-3213(01)00418-7}{\emph{Nucl. Phys.}
  {\bfseries B617} (2001) 441}
  [\href{https://arxiv.org/abs/hep-ph/0105011}{{\ttfamily hep-ph/0105011}}].

\bibitem{Hambye:2003cy}
T.~Hambye, S.~Peris and E.~de~Rafael, \emph{{Delta I = 1/2 and epsilon-prime /
  epsilon in large N(c) QCD}},
  \href{https://doi.org/10.1088/1126-6708/2003/05/027}{\emph{JHEP} {\bfseries
  05} (2003) 027} [\href{https://arxiv.org/abs/hep-ph/0305104}{{\ttfamily
  hep-ph/0305104}}].

\bibitem{Buras:2015xba}
A.~J. Buras and J.-M. G\'erard, \emph{{Upper bounds on
  $\varepsilon'/\varepsilon$ parameters B$_{6}^{(1/2)}$ and B$_{8}^{(3/2)}$
  from large N QCD and other news}},
  \href{https://doi.org/10.1007/JHEP12(2015)008}{\emph{JHEP} {\bfseries 12}
  (2015) 008} [\href{https://arxiv.org/abs/1507.06326}{{\ttfamily
  1507.06326}}].

\bibitem{Buras:2016fys}
A.~J. Buras and J.-M. Gerard, \emph{{Final state interactions in $K\rightarrow
  \pi \pi $ decays: $\Delta I=1/2$ rule vs. $\varepsilon '/\varepsilon $}},
  \href{https://doi.org/10.1140/epjc/s10052-016-4586-7}{\emph{Eur. Phys. J.}
  {\bfseries C77} (2017) 10}
  [\href{https://arxiv.org/abs/1603.05686}{{\ttfamily 1603.05686}}].

\bibitem{Gisbert:2017vvj}
H.~Gisbert and A.~Pich, \emph{{Direct CP violation in $K^0\to\pi\pi$: Standard
  Model Status}}, \href{https://doi.org/10.1088/1361-6633/aac18e}{\emph{Rept.
  Prog. Phys.} {\bfseries 81} (2018) 076201}
  [\href{https://arxiv.org/abs/1712.06147}{{\ttfamily 1712.06147}}].

\bibitem{Buras:2015yba}
A.~J. Buras, M.~Gorbahn, S.~J$\ddot{\textrm{a}}$ger and M.~Jamin,
  \emph{{Improved anatomy of $\varepsilon'/\varepsilon$ in the Standard
  Model}}, \href{https://doi.org/10.1007/JHEP11(2015)202}{\emph{JHEP}
  {\bfseries 11} (2015) 202}
  [\href{https://arxiv.org/abs/1507.06345}{{\ttfamily 1507.06345}}].

\bibitem{Kitahara:2016nld}
T.~Kitahara, U.~Nierste and P.~Tremper, \emph{{Singularity-free next-to-leading
  order $\Delta$S = 1 renormalization group evolution and
  $\epsilon_K'/\epsilon_K$ in the Standard Model and beyond}},
  \href{https://doi.org/10.1007/JHEP12(2016)078}{\emph{JHEP} {\bfseries 12}
  (2016) 078} [\href{https://arxiv.org/abs/1607.06727}{{\ttfamily
  1607.06727}}].

\bibitem{Gibbons:1993zq}
L.~K. Gibbons et~al., \emph{{Measurement of the CP violation parameter
  Re($\epsilon^{\prime} / \epsilon$)}},
  \href{https://doi.org/10.1103/PhysRevLett.70.1203}{\emph{Phys. Rev. Lett.}
  {\bfseries 70} (1993) 1203}.

\bibitem{Barr:1993rx}
{\scshape NA31} collaboration, G.~D. Barr et~al., \emph{{A New measurement of
  direct CP violation in the neutral kaon system}},
  \href{https://doi.org/10.1016/0370-2693(93)91599-I}{\emph{Phys. Lett.}
  {\bfseries B317} (1993) 233}.

\bibitem{Batley:2002gn}
{\scshape NA48} collaboration, J.~R. Batley et~al., \emph{{A Precision
  measurement of direct CP violation in the decay of neutral kaons into two
  pions}}, \href{https://doi.org/10.1016/S0370-2693(02)02476-0}{\emph{Phys.
  Lett.} {\bfseries B544} (2002) 97}
  [\href{https://arxiv.org/abs/hep-ex/0208009}{{\ttfamily hep-ex/0208009}}].

\bibitem{Abouzaid:2010ny}
{\scshape KTeV} collaboration, E.~Abouzaid et~al., \emph{{Precise Measurements
  of Direct CP Violation, CPT Symmetry, and Other Parameters in the Neutral
  Kaon System}}, \href{https://doi.org/10.1103/PhysRevD.83.092001}{\emph{Phys.
  Rev.} {\bfseries D83} (2011) 092001}
  [\href{https://arxiv.org/abs/1011.0127}{{\ttfamily 1011.0127}}].

\bibitem{Patrignani:2016xqp}
{\scshape Particle Data Group} collaboration, C.~Patrignani et~al.,
  \emph{{Review of Particle Physics}},
  \href{https://doi.org/10.1088/1674-1137/40/10/100001}{\emph{Chin. Phys.}
  {\bfseries C40} (2016) 100001}.

\bibitem{Cirigliano:2003nn}
V.~Cirigliano, A.~Pich, G.~Ecker and H.~Neufeld, \emph{{Isospin violation in
  epsilon-prime}},
  \href{https://doi.org/10.1103/PhysRevLett.91.162001}{\emph{Phys. Rev. Lett.}
  {\bfseries 91} (2003) 162001}
  [\href{https://arxiv.org/abs/hep-ph/0307030}{{\ttfamily hep-ph/0307030}}].

\bibitem{Lellouch:2000pv}
L.~Lellouch and M.~Luscher, \emph{{Weak transition matrix elements from finite
  volume correlation functions}},
  \href{https://doi.org/10.1007/s002200100410}{\emph{Commun. Math. Phys.}
  {\bfseries 219} (2001) 31}
  [\href{https://arxiv.org/abs/hep-lat/0003023}{{\ttfamily hep-lat/0003023}}].

\bibitem{Colangelo:2015kha}
G.~Colangelo, E.~Passemar and P.~Stoffer, \emph{{A dispersive treatment of
  $K_{\ell 4}$ decays}},
  \href{https://doi.org/10.1140/epjc/s10052-015-3357-1}{\emph{Eur. Phys. J.}
  {\bfseries C75} (2015) 172}
  [\href{https://arxiv.org/abs/1501.05627}{{\ttfamily 1501.05627}}].

\bibitem{lattice_prospect}
C.~Kelly, ``Progress in lattice in the kaon system.'' Talk at CKM 2018
  workshop, September, 2018.

\bibitem{Kitahara:2016otd}
T.~Kitahara, U.~Nierste and P.~Tremper, \emph{{Supersymmetric Explanation of CP
  Violation in $K\to \pi\pi$ Decays}},
  \href{https://doi.org/10.1103/PhysRevLett.117.091802}{\emph{Phys. Rev. Lett.}
  {\bfseries 117} (2016) 091802}
  [\href{https://arxiv.org/abs/1604.07400}{{\ttfamily 1604.07400}}].

\bibitem{Kagan:1999iq}
A.~L. Kagan and M.~Neubert, \emph{{Large Delta I = 3/2 contribution to
  epsilon-prime / epsilon in supersymmetry}},
  \href{https://doi.org/10.1103/PhysRevLett.83.4929}{\emph{Phys. Rev. Lett.}
  {\bfseries 83} (1999) 4929}
  [\href{https://arxiv.org/abs/hep-ph/9908404}{{\ttfamily hep-ph/9908404}}].

\bibitem{Grossman:1999av}
Y.~Grossman, M.~Neubert and A.~L. Kagan, \emph{{Trojan penguins and isospin
  violation in hadronic B decays}},
  \href{https://doi.org/10.1088/1126-6708/1999/10/029}{\emph{JHEP} {\bfseries
  10} (1999) 029} [\href{https://arxiv.org/abs/hep-ph/9909297}{{\ttfamily
  hep-ph/9909297}}].

\bibitem{Crivellin:2010ys}
A.~Crivellin and M.~Davidkov, \emph{{Do squarks have to be degenerate?
  Constraining the mass splitting with Kaon and D mixing}},
  \href{https://doi.org/10.1103/PhysRevD.81.095004}{\emph{Phys. Rev.}
  {\bfseries D81} (2010) 095004}
  [\href{https://arxiv.org/abs/1002.2653}{{\ttfamily 1002.2653}}].

\bibitem{Crivellin:2017gks}
A.~Crivellin, G.~D'Ambrosio, T.~Kitahara and U.~Nierste, \emph{{$K\to \pi
  \nu\overline{\nu}$ in the MSSM in light of the
  $\epsilon^{\prime}_K/\epsilon_K$ anomaly}},
  \href{https://doi.org/10.1103/PhysRevD.96.015023}{\emph{Phys. Rev.}
  {\bfseries D96} (2017) 015023}
  [\href{https://arxiv.org/abs/1703.05786}{{\ttfamily 1703.05786}}].

\bibitem{Buras:2014sba}
A.~J. Buras, F.~De~Fazio and J.~Girrbach, \emph{{$\Delta I=1/2$ rule,
  $\varepsilon '/\varepsilon $ and $K\rightarrow \pi \nu \bar{\nu }$ in $Z'
  (Z)$ and $G' $ models with FCNC quark couplings}},
  \href{https://doi.org/10.1140/epjc/s10052-014-2950-z}{\emph{Eur. Phys. J.}
  {\bfseries C74} (2014) 2950}
  [\href{https://arxiv.org/abs/1404.3824}{{\ttfamily 1404.3824}}].

\bibitem{Buras:2015yca}
A.~J. Buras, D.~Buttazzo and R.~Knegjens, \emph{{$ K\to \pi \nu \overline{\nu}
  $ and ε′/ε in simplified new physics models}},
  \href{https://doi.org/10.1007/JHEP11(2015)166}{\emph{JHEP} {\bfseries 11}
  (2015) 166} [\href{https://arxiv.org/abs/1507.08672}{{\ttfamily
  1507.08672}}].

\bibitem{Tanimoto:2016yfy}
M.~Tanimoto and K.~Yamamoto, \emph{{Probing SUSY with 10 TeV stop mass in rare
  decays and CP violation of kaon}},
  \href{https://doi.org/10.1093/ptep/ptw160}{\emph{PTEP} {\bfseries 2016}
  (2016) 123B02} [\href{https://arxiv.org/abs/1603.07960}{{\ttfamily
  1603.07960}}].

\bibitem{Endo:2016aws}
M.~Endo, S.~Mishima, D.~Ueda and K.~Yamamoto, \emph{{Chargino contributions in
  light of recent $\epsilon'/\epsilon$}},
  \href{https://doi.org/10.1016/j.physletb.2016.10.009}{\emph{Phys. Lett.}
  {\bfseries B762} (2016) 493}
  [\href{https://arxiv.org/abs/1608.01444}{{\ttfamily 1608.01444}}].

\bibitem{Endo:2017ums}
M.~Endo, T.~Goto, T.~Kitahara, S.~Mishima, D.~Ueda and K.~Yamamoto,
  \emph{{Gluino-mediated electroweak penguin with flavor-violating trilinear
  couplings}}, \href{https://doi.org/10.1007/JHEP04(2018)019}{\emph{JHEP}
  {\bfseries 04} (2018) 019}
  [\href{https://arxiv.org/abs/1712.04959}{{\ttfamily 1712.04959}}].

\bibitem{Endo:2016tnu}
M.~Endo, T.~Kitahara, S.~Mishima and K.~Yamamoto, \emph{{Revisiting Kaon
  Physics in General $Z$ Scenario}},
  \href{https://doi.org/10.1016/j.physletb.2017.05.026}{\emph{Phys. Lett.}
  {\bfseries B771} (2017) 37}
  [\href{https://arxiv.org/abs/1612.08839}{{\ttfamily 1612.08839}}].

\bibitem{Bobeth:2017xry}
C.~Bobeth, A.~J. Buras, A.~Celis and M.~Jung, \emph{{Yukawa enhancement of
  $Z$-mediated new physics in $\Delta S = 2$ and $\Delta B = 2$ processes}},
  \href{https://doi.org/10.1007/JHEP07(2017)124}{\emph{JHEP} {\bfseries 07}
  (2017) 124} [\href{https://arxiv.org/abs/1703.04753}{{\ttfamily
  1703.04753}}].

\bibitem{NA62:2017rwk}
{\scshape NA62} collaboration, E.~Cortina~Gil et~al., \emph{{The Beam and
  detector of the NA62 experiment at CERN}},
  \href{https://doi.org/10.1088/1748-0221/12/05/P05025}{\emph{JINST} {\bfseries
  12} (2017) P05025} [\href{https://arxiv.org/abs/1703.08501}{{\ttfamily
  1703.08501}}].

\bibitem{DAmbrosio:2017klp}
G.~D'Ambrosio and T.~Kitahara, \emph{{Direct $CP$ Violation in $K \to \mu^+
  \mu^-$}}, \href{https://doi.org/10.1103/PhysRevLett.119.201802}{\emph{Phys.
  Rev. Lett.} {\bfseries 119} (2017) 201802}
  [\href{https://arxiv.org/abs/1707.06999}{{\ttfamily 1707.06999}}].

\bibitem{Chobanova:2017rkj}
V.~Chobanova, G.~D'Ambrosio, T.~Kitahara, M.~Lucio~Martinez,
  D.~Martinez~Santos, I.~S. Fernandez et~al., \emph{{Probing SUSY effects in
  $K_S^0\rightarrow\mu^+\mu^-$}},
  \href{https://doi.org/10.1007/JHEP05(2018)024}{\emph{JHEP} {\bfseries 05}
  (2018) 024} [\href{https://arxiv.org/abs/1711.11030}{{\ttfamily
  1711.11030}}].

\end{thebibliography}\endgroup

\end{document}